\documentclass[1p]{elsarticle}
\usepackage{amsmath,amssymb,graphicx,hyperref}
\hypersetup{colorlinks=true}
\begin{document}

\title{Dynamical system analysis of quantum tunneling in an asymmetric double-well potential}
\author[inst1]{Swetamber Das\corref{cor1}}
\ead{swetamber.p@srmap.edu.in}
\author[inst1]{Arghya Dutta\corref{cor1}}
\ead{arghya.d@srmap.edu.in}
\affiliation[inst1]{organization={Department of Physics, SRM University-AP}, city={Amaravati}, postcode={522 240}, state={Andhra Pradesh}, country={India}}
\cortext[cor1]{Corresponding author}

\begin{abstract}
We study quantum tunneling in an asymmetric double-well potential using a dynamical systems--based approach rooted in the Ehrenfest formalism. In this framework, the time evolution of a Gaussian wave packet is governed by a hierarchy of coupled equations linking lower- and higher-order position moments. An approximate closure scheme, required to render the system tractable, yields a reduced dynamical system for the mean and variance, with skewness explicitly entering due to the potential’s asymmetry.  Stability analysis of this system identifies energy thresholds for detectable tunneling across the barrier and reveals regimes where tunneling, though theoretically allowed, remains practically undetectable. Comparison with full numerical solutions of the time-dependent Schr\"odinger equation shows that, beyond reproducing key tunneling features, the dynamical systems approach provides an interpretable description of quantum transport through tunneling in an effective asymmetric two-level system.
\end{abstract}

\maketitle

\section{Introduction}

Quantum mechanical tunneling in double-well potentials remains a central framework for probing the dynamics of two-level quantum systems \cite{leggett1987dynamics, grifoni1996dynamics, halataei2017tunnel}. Examples of tunneling in effective two-level systems arise in many microscopic and molecular processes: inversion of ammonia \cite{swalen1962potentiala}; double hydrogen tunneling in porphycene \cite{mengesha2013vibrations}; electron tunneling between two coupled quantum dots \cite{burkard1999coupled}; and proton tunneling in malonaldehyde \cite{yagi2001generation, firth1991tunable, ghosh2015optimised}, partially deuterated malonaldehyde \cite{baughcum1981microwave, jahr2020instanton}, and vinyl radicals \cite{tanaka2004determination}. Similar tunneling phenomena also occur in macroscopic quantum systems exhibiting effective two-level behavior, for instance in Bose--Einstein condensates \cite{albiez2005direct, shin2005optical, schumm2005matterwave, hall2007condensate} and in superconducting quantum interference devices \cite{devoret1985measurements, makhlin2001quantumstate, levy2007ac, clarke2008superconducting, lisenfeld2015observation}. Remarkably, despite the complexities of macroscopic tunneling involving many particles, the dynamics of such systems can often be modeled with reasonable accuracy as a single particle moving and tunneling in an effective double-well potential $\phi(x)$ defined over some collective macroscopic coordinate $x$ \cite{rastelli2012semiclassical}.

However, studying the full quantum dynamics of even a simplified single-particle system becomes challenging when the potential cannot be treated perturbatively, thereby falling outside the scope of Fermi’s golden rule \cite{cohenTannoudji2020quantuma}. In such cases---for instance tunneling in double-well potentials---non-perturbative instanton-based approaches have been used \cite{paranjape2022theory, richardson2011ringpolymer, cvitas2016locating, jahr2020instanton}. In addition, the semiclassical Wentzel--Kramers--Brillouin (WKB) approximation has been widely employed, often together with the simplifying assumption that the two wells can be modeled as simple harmonic potentials \cite{garg2000tunnel, song2008tunneling, rastelli2012semiclassical, song2015localization, halataei2017tunnel}. Despite its broad utility, however, the WKB approach faces some limitations.

A central restriction of the WKB approximation arises from its validity condition, which requires the particle momentum $p(x)=\sqrt{2m|E-\phi(x)|}$ to remain sufficiently large. This assumption breaks down near classical turning points, where the momentum vanishes, and within potential wells for low-energy states, particularly the ground state, where both kinetic and potential energies may become small \cite{halataei2017tunnel}. Further, especially near the boundaries where wave functions must be matched across classically allowed and forbidden regions, the standard WKB wavefunctions often require fine-tuning \cite{wine2025modified}.

In addition, traditionally, the WKB approach has been mainly employed to obtain approximate solutions of the \emph{time-independent} Schr\"odinger equation. In contrast, time-dependent WKB and other such semiclassical methods remain an active area of research \cite{huber1987generalized, huber1988generalized, boiron1998complex, goldfarb2008complex, chou2008quantum}. This limitation becomes particularly significant when studying tunneling---an inherently time-dependent quantum phenomenon \cite{bender2011quantum}. For instance, in a double-well potential, tunneling is theoretically permitted at all sub-barrier energies. While several studies have addressed quantum tunneling detection and measurement in a double-well potential using open-system approaches \cite{gisin1993quantum, altenmuller1994quantum, xiao2012coherent, reynoso2023quantum}, tunneling becomes practically undetectable for states deep within the wells since it is exponentially suppressed. This poses a significant impediment in studying systems like ultracold optical lattices \cite{Gross_Bloch_2017}, where such suppression prohibits the detection of small tunnel splittings \cite{Raso2025nonadiabatic}. Overcoming this challenge to predict a practical quantitative energy threshold for detectable quantum transport through tunneling requires a theoretical framework capable of modeling time-dependent quantum evolution directly, a task for which the standard time-dependent WKB approximation is ill-suited.

To bridge this gap, we turn to alternative complementary theoretical frameworks that naturally connect classical and quantum dynamics. Among them, a systematic approach is provided by the Ehrenfest theorem \cite{ehrenfest1927bemerkung, heller1975timedependent, pattanayak1994semiquantal}, which, although often introduced as a means to extract the semiclassical limit, yields exact quantum mechanical equations governing the time evolution of expectation values of operators, and is therefore not limited to describing the classical limit of quantum mechanics \cite{roy2001chaos, biswas2018propagation, chawla2019quantum, sarkar2020nonlinear, bhattacharyya2021new}. For a quantum operator $\mathcal{O}$ that does not depend explicitly on time, the time evolution of its expectation value in a system described by a Hamiltonian $H$ is given by the Ehrenfest theorem:
\begin{align}
\label{eq:ehrenfest}
\frac{\mathrm{d}}{\mathrm{d}t}\langle \mathcal{O} \rangle = \frac{1}{i\hbar}\langle [\mathcal{O},H]\rangle,
\end{align}
which is fundamentally equivalent to the Heisenberg equation of motion for operators and provides a general framework for extracting quantum dynamics directly from the evolution of expectation values.

Recently, the Ehrenfest formalism has been employed to model tunneling dynamics in symmetric double-well \cite{Choi2015, ray2025dynamical} and Morse \cite{sarkar2024quantum} potentials, treating them as effective dynamical systems. For a single particle initially described by a Gaussian wave packet, Eq.~\ref{eq:ehrenfest} enables the reconstruction of the full time evolution of the system through the dynamics of its moments $\langle x^n\rangle$, where $n$ denotes the order of the moment. While for Hamiltonians containing terms beyond quadratic order, the lower-order moment equations couple to higher-order ones, necessitating approximate closures \cite{Sundaram1995, brizuela2014statistical}, this formalism naturally highlights the distinction between classical and quantum dynamics, since the presence of uncertainty relations among non-commuting observables makes the quantum moments evolve very differently from their classical counterparts \cite{baird1972moments, styer1990motion, ballentine1998moment, brizuela2014classical, brizuela2014statistical}. 

Here we employ the dynamical systems approach to study quantum tunneling of a particle, represented by a Gaussian wave packet, in a one-dimensional asymmetric double-well potential. Such asymmetry breaks the system's reflection symmetry, inducing highly directional tunneling important in several physical processes, including tunneling in strong-field ionization \cite{Jia2022} and proton transfer in photosynthesis \cite{Marais2018future}. Through stability analysis of a suitably reduced four-dimensional dynamical system governing the wave packet's mean and variance, we compute a practical estimate of the energy threshold required for a detectable signature of tunneling across the barrier between the two wells. This threshold is determined by the skewness of the wave packet's position distribution, which quantifies the potential's asymmetry. Consequently, we incorporate skewness as a key ingredient in our analysis to determine the energy range above which detectable tunneling occurs, indicated by switching of the mean position $\langle x \rangle$ between the two wells as the system evolves. Comparison with numerical solutions to the full time-dependent Schr\"odinger equation demonstrates that the reduced dynamical model reliably captures the essential features of asymmetric tunneling dynamics, while offering additional insight on practical tunneling energy thresholds.

\section{Method}

\subsection{Double-well potentials}

\begin{figure}[!htbp]
\centering
\includegraphics{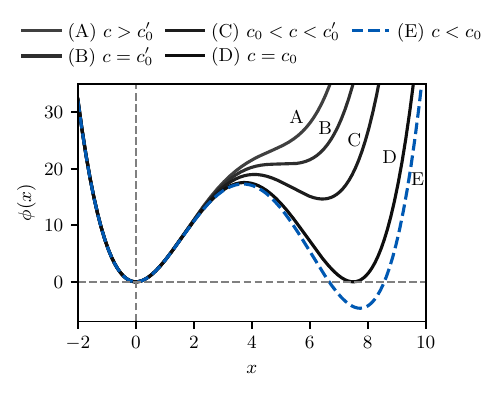}
\caption{\textbf{Symmetric and asymmetric double-well potentials.} Representative forms of the quartic potential $\phi(x)$ (Eq.~\ref{eq:potential}) are shown with fixed $a=10$ and $b=4$, and varying $c$. Curves A--E illustrate the transition from a single well to symmetric and asymmetric double-well configurations, respectively, as $c$ decreases. The asymmetric double-well potential with $c=0.35$ (curve E, blue, dashed) is studied in this work.}
\label{fig:potential}
\end{figure}

We begin with a brief discussion on the double-well potentials which can be represented by 
\begin{align}
\phi(x) = \frac{a}{2} x^2 - \frac{b}{3} x^3 + \frac{c}{4}x^4,
\label{eq:potential}
\end{align}
where $a$, $b$, and $c$ are positive parameters. Depending on the choice of these parameters, Eq.~\ref{eq:potential} can represent either a symmetric or an asymmetric double-well potential. The potential $\phi(x)$ reduces to the symmetric potential studied by Ray et al.~\cite{ray2025dynamical}, $U(x) = \omega^2 x^2 - \omega\sqrt{\lambda}\,x^3 + \lambda x^4/4 $, under the parameter mapping $a = 2\omega^2$, $b = 3\sqrt{\lambda}\,\omega$, and $c=\lambda$.

The three stationary points of the asymmetric potential $\phi(x)$ are obtained by solving $\phi'(x)=0$, which gives $x = 0$ and $x=\beta_\pm$ where $\beta_\pm \equiv (b\pm \sqrt{b^2-4ac})/2c$. The solutions $\beta_\pm$ are real only if the discriminant is non-negative, or equivalently $c \leq c_0' \equiv b^2/4a$. The two stationary points at $\beta_\pm$ coalesce for $c = c_0'$, giving a single stationary point $\beta \equiv b/2c$. Evaluating $\phi''(x)$ at the stationary point $x=0$ reveals $\phi''(0)=a$, showing that $x=0$ is a local maximum if $a<0$ and is a local minimum if $a>0$. Throughout this work, we will take $a>0$.

At $c=c_0'$, the stationary point at $x=\beta \equiv b/2c$ satisfies $\phi''(\beta)=0$ (and $\phi'''(\beta)=b \neq 0$), corresponding to the appearance of an inflection point. For $c > c_0'$, the potential has only one minimum at $x=0$.

To obtain a symmetric double well, we additionally require $\phi(x)=0$ to have nontrivial roots. Solving for $\phi(x)=0$ yields $x = 0$ and $x = \alpha_\pm$ where $\alpha_\pm= 2b/3c \pm (2/c)\sqrt{b^2/9-ac/2}$. When $c>c_0\equiv 2b^2/9a$, $x=0$ is the only real root. At $c=c_0$, the two roots $\alpha_\pm$ merge to a single value $\alpha \equiv 2b/3c = 3a/b$ (since $c=2b^2/9a$), where $\phi'(\alpha)=0$ and $\phi''(\alpha)=a$, giving a minimum of the potential since we took $a>0$ for this paper. Again, notice that the non-zero roots of $\phi'(x)=0$ at $c=c_0\equiv 2b^2/9a$ take the values $\beta_+ = 3a/b = \alpha$, corresponding to the second well at finite $x$, and $\beta_-=3a/2b$, corresponding to the potential hill located between the two minima.

Figure~\ref{fig:potential} summarizes the discussion. For positive fixed values of $a$ and $b$, the potential has a single minimum at $x=0$ when $c>c_0' (\equiv b^2/4a)$ (curve A). An inflection point appears when $c=c_0'$ (curve B). For $c_0<c<c_0'$ with $c_0\equiv 2b^2/9a$, the potential contains a global minimum at $x=0$ and a local minimum at $x=\beta_+$ (curve C), separated by a barrier at $x=\beta_-$. At $c=c_0$, the two minima become degenerate, yielding a symmetric double well (curve D). Finally, for $c<c_0$, the global minimum shifts to $x=\beta_+$, where $\phi(\beta_+)<\phi(0)=0$, giving an asymmetric double well (curve E). In this paper, we focus on the asymmetric double-well potential described by curve E. The specific parameter values $a=10$, $b=4$, and $c=0.35$ used in this work are chosen to realize a possible asymmetric double-well configuration---the choice is not unique. That is, reducing $a$ (or increasing $b$, which enters the potential with a negative sign) at fixed $c$ has a qualitative effect similar to decreasing $c$ at fixed $a$ and $b$ (curves A--E in Fig.~\ref{fig:potential}). As long as the asymmetric shape of the double-well is preserved, the tunneling dynamics are expected to remain qualitatively unchanged.

\subsection{Dynamical equations}

Consider a particle of mass $m=1$ moving in the asymmetric double-well potential described by Eq.~\ref{eq:potential}. The Hamiltonian of the system is
\begin{align}
H = \frac{p^2}{2} + \phi(x) = \frac{p^2}{2} + \frac{a}{2} x^2 - \frac{b}{3} x^3 + \frac{c}{4} x^4.
\label{eq:hamiltonian}
\end{align}

To construct the dynamical system comprising time evolution of the mean position $\langle x \rangle$ and the variance $V\equiv \langle x^2 \rangle - \langle x \rangle^2$ of the particle, we employ the Ehrenfest theorem (Eq.~\ref{eq:ehrenfest}) and simplify the commutators to obtain
\begin{align}
\frac{\mathrm{d} \langle x \rangle }{\mathrm{d}t} 
& = \langle p \rangle ,\label{eq:dxdt} \\
\frac{\mathrm{d}^2 \langle x \rangle }{\mathrm{d}t^2} 
& = - \langle  \phi' \rangle \nonumber \\
& = - a \langle x \rangle
+ b \left[ V+ \langle x \rangle ^2 \right]
- c \left[ S + 3 V \langle x \rangle + \langle x \rangle^3 \right] ,\label{eq:d2xdt2} \\
\frac{\mathrm{d}V}{\mathrm{d}t} 
& = \langle xp+px \rangle - 2\langle x \rangle \langle p \rangle, \label{eq:dvdt} \\
\frac{\mathrm{d}^2 V}{\mathrm{d}t^2} & = 2V_p - 2 (\langle x \phi' \rangle - \langle x\rangle \langle  \phi' \rangle)  \nonumber \\
& =
4E - 2\langle p \rangle^2 
- a \left [ 4V + 2 \langle x \rangle^2 \right] \nonumber\\
& \quad + b \left[ \frac{10}{3}S + 8 V \langle x \rangle + \frac{4}{3}\langle x \rangle^3 \right] \nonumber\\
& \quad - c \left[ 9V^2 + 10 S \langle x \rangle + 12 V \langle x \rangle^2 + \langle x \rangle^4 \right]\label{eq:d2vdt2}.
\end{align}

In Eq.~\ref{eq:d2vdt2}, $V_p \equiv \langle p^2 \rangle - \langle p \rangle ^2$ is the variance of the particle's momentum. It is related to the total energy of the system $E (\equiv \langle H \rangle)$ since $\langle p^2 \rangle = 2E - 2 \langle \phi \rangle$. The mean energy $E$ is the only conserved quantity of this system, serving as a control parameter for the dynamics. In Eqs.~\ref{eq:d2xdt2} and \ref{eq:d2vdt2}, $S \equiv \langle (x - \langle x \rangle)^3\rangle$ is the skewness and $K \equiv \langle (x - \langle x \rangle)^4\rangle$ is the kurtosis, which is simplified with the Gaussian approximation $K=3V^2$. This approximation is reasonable because we represent the unit-mass particle as a Gaussian wave packet
\begin{equation}
\label{eq:wavepacket}
\psi(x,t) = \frac{1}{\sqrt[4]{2\pi V}} \exp\left[{-\frac{(x-\langle x\rangle)^2}{4V}} + ik(x-\langle x \rangle)\right],
\end{equation}
where $k$ is the wave number, and expect its shape to retain the properties of a Gaussian distribution, to a large extent. Notice that both the variance and the mean position of this wave packet depend on time.

If we start with a Gaussian wave packet initialized at $\langle x \rangle(0) = 0$, variance $V(0)=V_0$, and wave number $k(0)=k_0$ at time $t=0$, computing the expectation value of the Hamiltonian (Eq.~\ref{eq:hamiltonian}) in this wave packet provides a connection between the conserved total energy $E$ and the initial variance $V_0$:
\begin{equation}
\label{eq:initial_energy}
E=\frac{1}{8V_0} + \frac{k_0^2}{2} + \frac{a}{2}V_0 + \frac{3}{4}cV_0^2,
\end{equation}
where we have set $\hbar=1$. With these definitions, we now proceed to analyze the four-dimensional dynamical system.

\section{Results and Discussions}
\subsection{Stability analysis}
\begin{figure*}[!htbp]
\centering
\includegraphics[width=\textwidth]{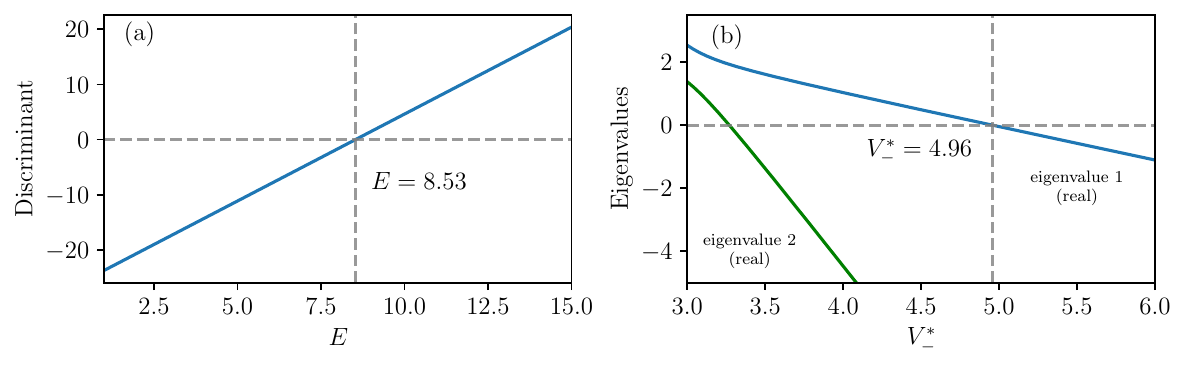}
\caption{\textbf{Stability analysis.} (a) The solutions for $V_-^\ast$ of the quadratic equation (Eq.~\ref{eq:V_ast_quadratic}) are real only when the discriminant is non-negative. This condition is satisfied for energies $E \geq 8.53$, which is the minimum energy needed for the existence of the fixed point $V_-^\ast$. (b) The real parts of the two eigenvalues of the stability matrix for the fixed point $(\beta_-, 0, V_-^\ast, 0)$, associated with the potential hill, are plotted against $V_-^\ast$. Both eigenvalues are negative beyond $V^\ast_- = 4.96$, corresponding to $E = 10.60$. This signifies the transition of the potential hill from a classically unstable fixed point to a stable one, indicating the onset of tunneling.
}
\label{fig:stability}
\end{figure*}

To analyze the stability of the four-dimensional dynamical system described by Eqs.~\ref{eq:dxdt}--\ref{eq:d2vdt2}, we first linearize them as follows:
\begin{align}
\frac{\mathrm{d}^2}{\mathrm{d}t^2} \begin{pmatrix} \delta \langle x \rangle \\ \delta V \end{pmatrix} = \mathbf{A} \begin{pmatrix} \delta \langle x \rangle \\ \delta V \end{pmatrix},
\end{align}
where the stability matrix $\mathbf{A}$ is given by:
\begin{align}\label{eq:matrix_A}
\mathbf{A} = \begin{pmatrix}
A_{11} & A_{12}\\
A_{21} & A_{22}
\end{pmatrix}.
\end{align}
Here
\begin{align*}
A_{11} &= -a + 2b \langle x \rangle - 3c \langle x \rangle^2 - 3c V \\   
A_{12} &= b - 3c \langle x \rangle \\
A_{21} &= -4a \langle x \rangle + 4b \langle x \rangle^2 - 4c \langle x \rangle^3 + 8b V - 24c \langle x \rangle V - 10cS \\
A_{22} &= -4a + 8b \langle x \rangle - 18c V - 12c \langle x \rangle^2.
\end{align*}

The fixed points of the system are of the form $(\langle x\rangle^\ast, 0, V^\ast, 0)$. For such a fixed point to exist, skewness $S$ must be set by $V^\ast$. The constant value of skewness at the fixed point is a consequence of reducing the dynamical system to four dimensions involving mean and variance (with their time derivatives). We find the relation between them by equating Eq.~\ref{eq:d2xdt2} to zero (and since $c \neq 0$):
\begin{align} \label{eq:Skewness_relation}
S = -\frac{1}{c} \left[ a \langle x \rangle^\ast - b V^\ast - b \langle x \rangle^{\ast 2} + 3c V^\ast \langle x \rangle^\ast + c \langle x \rangle^{\ast 3} \right].
\end{align}
We substitute this expression for $S$ in Eq.~\ref{eq:d2vdt2}  to obtain an equation for $V^\ast$:
\begin{align} \label{eq:V_ast_quadratic}
& \frac{9c}{4} V^{*2} + \left( a - \frac{5b^2}{6c} + 3 b \langle x \rangle^* - \frac{9 c \langle x \rangle^{*2}}{2} \right) V^* + \nonumber                                                          \\
& \left[ \frac{5 ab \langle x \rangle^*}{6 c} - 2a \langle x \rangle^{*2} - \frac{5 b^2 \langle x \rangle^{*2}}{6 c} + 3 b \langle x \rangle^{*3} - \frac{9c \langle x \rangle^{*4}}{4} \right] = E.
\end{align}
This quadratic equation can be solved numerically for given values of $\langle x\rangle^\ast$ and energy $E$. Real solutions for $V^\ast$ exist only when the discriminant of Eq.~\ref{eq:V_ast_quadratic} is non-negative.

Next, we analyze the existence and stability of the three fixed points given by (a) $(0, 0, V^\ast_0, 0)$, (b)  $(\beta_-, 0, V_-^\ast,0)$, and (c) $(\beta_+, 0, V_+^\ast,0)$. The fixed points (a) and (c) correspond to the two minima and (b) indicates the local maxima of the potential. As mentioned in Fig.~\ref{fig:potential}, the parameters of the asymmetric double-well potential (Eq.~\ref{eq:potential}) used in this study are $a=10$, $b=4$, and $c=0.35$. The fixed points discussed here arise in the reduced dynamical system defined by Eqs.~\ref{eq:dxdt}--\ref{eq:d2vdt2}. They represent the stationary points of its four-dimensional state space, and do not correspond to physical quantum states.

We first consider the case of the local maxima at $x = \beta_- = (b- \sqrt{b^2-4ac})/2c $, which is about 3.69 for the chosen parameter values and numerically analyze its stability. The discriminant from Eq.~\ref{eq:V_ast_quadratic} as a function of energy is plotted in Fig.~\ref{fig:stability}(a), and it shows that real solutions exist when $E\geq 8.53$. Stability analysis further shows that the fixed point  $(\beta_-, 0, V_-^\ast,0)$ becomes stable when the real parts of both eigenvalues of the matrix $\mathbf{A}$ (Eq.~\ref{eq:matrix_A}) are negative. We compute and plot them in Fig.~\ref{fig:stability}(b), which shows that both are negative for $V^\ast_-\geq 4.96$. This value of $V^\ast_-$ corresponds to $E = 10.60$ from Eq.~\ref{eq:V_ast_quadratic}. This is the energy threshold above which we expect tunneling to occur. We now list our findings below:
\begin{enumerate}
\item Real solutions for $V_-^\ast$ do not exist when  $E < 8.53$. There are no fixed points below this energy threshold.
\item In the regime $8.53\leq E < 10.60$, the fixed point exists, but it is unstable and tunneling does not occur.
\item For $10.60 \leq E < 17.31$, the classically unstable fixed point at the potential hill top becomes stable and tunneling takes place.
\item Energies higher than $E =17.31$ are irrelevant for tunneling because they exceed the barrier height.
\end{enumerate}

For the two potential minima, the corresponding fixed points (a) and (c) always exist for any positive value of $E$ and both are stable. To see this, we first consider the fixed point (a) for which $\langle x\rangle^\ast=0$. This yields the following condition for real solutions of $V_0^\ast$ to exist: $E \geq -( a - 5b^2/6c)^2/9c.$  Because $c > 0$, the lower bound on the inequality is trivially satisfied as $E$ must be non-negative. The other fixed point (c) $(\beta_+, 0, V_+^\ast,0)$ is located at the global minima on the right at $x = \beta_+ = 7.73$. Like the fixed point (a), we get a real solution for $V_+^\ast$ for any positive value of energy. The two fixed points (a) and (c) are therefore irrelevant for the tunneling dynamics. 

\subsection{Numerical Results: Dynamical systems approach}

\begin{figure*}[!htbp]
\includegraphics[width=\textwidth]{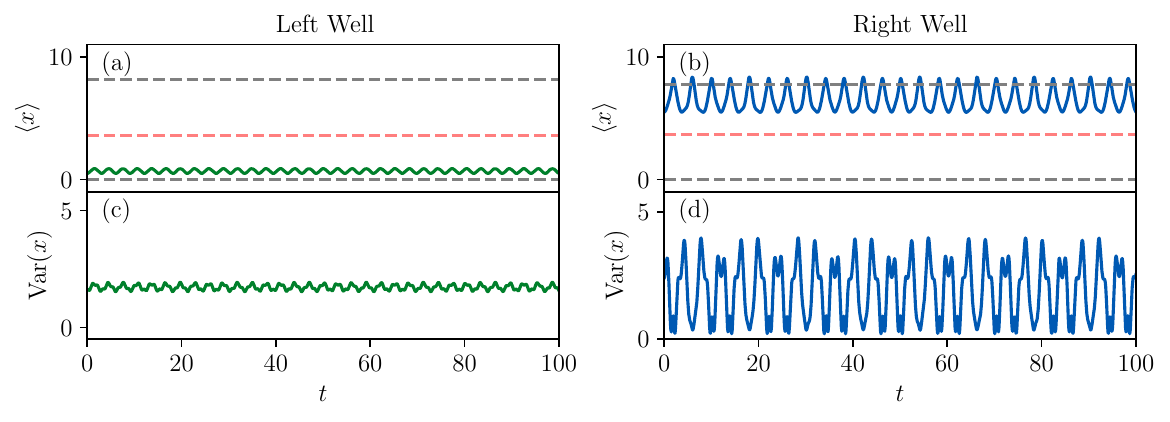}

\caption{\textbf{No tunneling (dynamical systems analysis).} The time evolution of the mean position $\langle x \rangle$ of the wave packet is shown for initializations in (a) the left well ($\langle x\rangle (0) = 0.5$, $E= 9.0$) and (b) the right well ($\langle x\rangle (0) = 5.50$, $E = \Delta + 9.0$, where $\Delta = 4.68$ is the energy difference between the two minima). In both cases, $\langle x \rangle$ oscillates but remains confined within its well. The red dashed line at $\langle x\rangle = 3.69$ indicates the maxima at the potential hill, and the gray dashed line at $\langle x\rangle = 0$ and $7.73$ mark the potential minima for the left and right wells, respectively. The variance in the right well (panel (d)) is larger than in the left well (panel (c)). This difference can be attributed to the higher energy ($E = \Delta + 9.0$) of the wave packet in the right well compared to the left well ($E = 9.0$).}
\label{fig:dyn_sys_no_tunnel-E-9}
\end{figure*}

\begin{figure*}[!htbp]
\includegraphics[width=\textwidth]{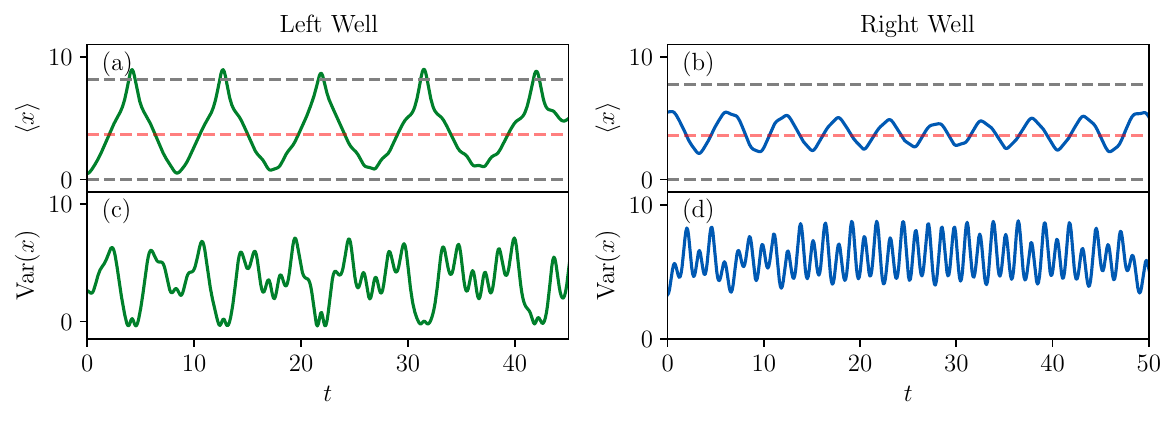}

\caption{\textbf{Tunneling (dynamical systems analysis).} The time evolution of the mean position $\langle x \rangle$ of the wave packet is shown for initializations in (a) the left well ($\langle x\rangle (0) = 0.5$, $E= 14.95$) and (b) the right well ($\langle x\rangle (0) = 5.50$, $E = \Delta + 14.95$, where $\Delta = 4.68$ is the energy difference between the two minima). In panels (a) and (b), $\langle x \rangle$ switches between the two wells---indicating detectable tunneling. The red dashed line at $\langle x\rangle = 3.69$ indicates the maxima at the potential hill, and the gray dashed line at $\langle x\rangle = 0$ and $7.73$ mark the potential minima for the left and right wells, respectively. The variance in the right well (panel (d)) is larger than in the left well (panel (c)), owing to the higher total energy. Interestingly, the variance in the right well oscillates rapidly, along with frequent barrier crossings, compared to the left well where the particle makes longer excursions between the wells.}
\label{fig:dyn_sys_tunnel-E-14-95}
\end{figure*}

Assuming the fixed point $(\beta_-, 0, V_-^\ast, 0)$ exists at the local maximum at $x = \beta_- = 3.69$, we solve the system of Eqs.~\ref{eq:dxdt}--\ref{eq:d2vdt2}. The skewness $S$ in these equations is calculated from Eq.~\ref{eq:Skewness_relation} for this fixed point. The initial conditions are chosen such that both the initial variance $V(0)=V_0$ and its time derivative $\mathrm{d}V/\mathrm{d}t|_{t=0}$ are zero.

We investigate the system's tunneling behavior below the energy threshold. The system is initialized within the left well at $\langle x\rangle(0) = 0.5$ with an energy $E = 9.0$, which is below the tunneling threshold of 10.60. The total energy and the initial wave vector $k_0$ (or equivalently the momentum $p_0$) determine the initial variance $V(0)$ of the wave packet via Eq.~\ref{eq:initial_energy}, allowing the dynamical equations to be numerically integrated. We set $k_0$ to zero throughout this work. The resulting time series for the mean position $\langle x\rangle$ (green curve, Fig.~\ref{fig:dyn_sys_no_tunnel-E-9}(a)) shows oscillations confined to the left well. In all subsequent $\langle x \rangle$ time series plots, the red dashed line marks the local maxima (barrier top), and gray dashed lines mark the left and right potential minima.

Similarly, when initialized in the right well with $\langle x\rangle(0) = 5.5$ (blue curve, Fig.~\ref{fig:dyn_sys_no_tunnel-E-9}(b)), the particle remains localized. In both cases, the absence of switching between the wells confirms that tunneling does not occur.  This localization is corroborated by the time evolution of the variance $V(t)$ which remains low, indicating no barrier crossings (Fig.~\ref{fig:dyn_sys_no_tunnel-E-9}(c) and (d)). Interestingly, fluctuations are smaller in the shallower left well than in the deeper right well. The larger fluctuations around the mean position in the right well suggest that tunneling is impending at this energy but has not yet been triggered.

At an energy of $E = 14.95$, which exceeds the tunneling threshold, we observe clear tunneling from the left to the right well. This is evidenced by the time series of $\langle x \rangle$ (green curve, Fig.~\ref{fig:dyn_sys_tunnel-E-14-95}(a)), which exhibits large-amplitude oscillations that frequently cross the potential barrier, indicating switching between wells. To observe the reverse process, i.e. tunneling from the right to the left well, the system requires additional energy equal to the energy difference between the two wells $\Delta = 4.68$. When initialized in the right well at a higher energy $E = 14.95 + \Delta$, the mean position (blue curve, Fig.~\ref{fig:dyn_sys_tunnel-E-14-95}(b)) shows similar switching dynamics, confirming continuous bi-directional tunneling. The mean variance during the left-to-right tunneling (Fig.~\ref{fig:dyn_sys_tunnel-E-14-95}(c), green curve) is smaller than its value during right-to-left tunneling (Fig.~\ref{fig:dyn_sys_tunnel-E-14-95}(d), blue curve). Interestingly, when initiated in the right well, the wave packet keeps switching rapidly across the barrier, but when initiated in the left well, it makes more sustained excursions into both wells, spending less time near the barrier.

\subsection{Numerical Results: Schr\"odinger Equation solution}

\begin{figure*}[!htbp]
\includegraphics[width=\textwidth]{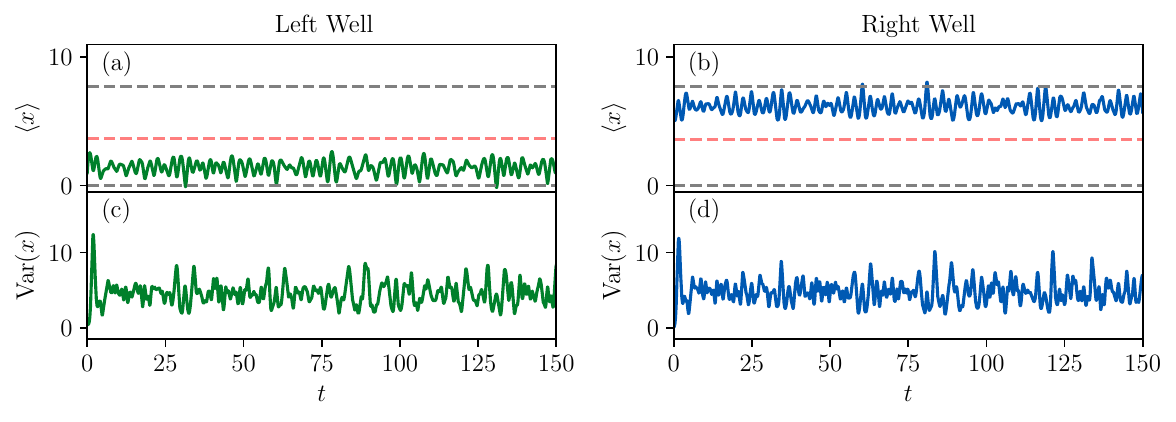}

\caption{\textbf{No tunneling (full Schr\"odinger simulation)}. Time evolution of the mean position $\langle x\rangle$ of the wave packet initialized at (a) $\langle x\rangle(0)=0.5$ and (b) $\langle x\rangle(0)=5.50$ in the left and right wells, respectively, with total energy $E=9.0+\Delta$ in both cases, where $\Delta=4.68$ is the energy offset given by the depth of the right well relative to the left. The wave packet oscillates within each well without crossing the barrier. The red dashed line ($\langle x\rangle=3.69$) indicates the barrier top, while the gray dashed lines at $\langle x\rangle=0$ and $7.73$ mark the locations of the left and right well minima, respectively. Panels (c) and (d) show the time evolution of the variance in the left and right wells. While the mean positions remain confined to their respective wells, the magnitude and temporal fluctuations of the variance are comparable in both cases.}
\label{fig:Schrodinger_no_tunnel-E-9}
\end{figure*}

\begin{figure*}[!htbp]
\includegraphics[width=\textwidth]{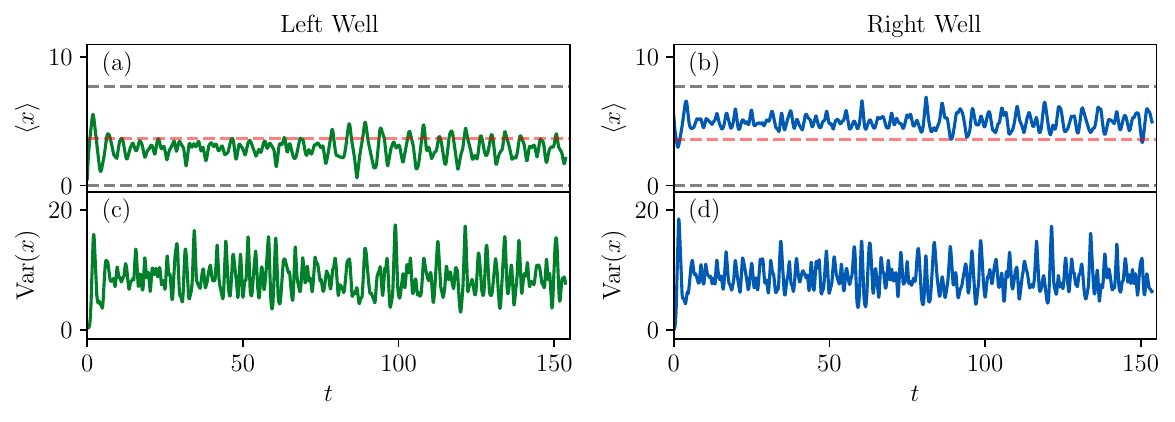}

\caption{\textbf{Tunneling (full Schr\"odinger simulation)}. Time evolution of the mean position $\langle x\rangle$ of the wave packet initialized at (a) $\langle x\rangle(0)=0.5$ and (b) $\langle x\rangle(0)=5.50$ in the left and right wells, respectively, with total energy $E=14.95+\Delta$ in both cases, where $\Delta=4.68$ is the energy offset given by the depth of the right well relative to the left. The mean position oscillates and frequently crosses the barrier, indicating detectable tunneling. The red dashed line ($\langle x\rangle=3.69$) indicates the barrier top, while the gray dashed lines at $\langle x\rangle=0$ and $7.73$ mark the locations of the left and right well minima, respectively. Panels (c) and (d) show the time evolution of the variance in the left and right wells. The variance is similar in both cases, as the time series of $\langle x\rangle$ for the two wells are only slightly shifted relative to one another.}
\label{fig:Schrodinger_tunnel-E-14-95}
\end{figure*}

We now compare the dynamical systems--based approach discussed in the previous section with a high-fidelity numerical solution of the time-dependent Schr\"odinger equation. The numerical integration is performed using the Crank--Nicolson method \cite{IzaacWang2018computational}, chosen for its unconditional stability and property of unitary evolution, which keeps the wave function's norm and the system's total energy nearly constant. To achieve high precision, we confine the particle in a box ($x \in [-100, 100]$) and then discretize it into a grid of $N = 10^5$ points ($\Delta x = 0.002$), while using a small time step of $\Delta t = 0.01$. Dirichlet boundary conditions ($\psi(x_{\min}) = \psi(x_{\max}) = 0$) are enforced at each time step. The large box size, compared to the separations between the potential extrema, ensures that the reflecting boundaries are located far from the region of interest, preventing any spurious reflections from affecting the dynamics over the simulation time. This implementation ensures that the drift in both the norm and the total energy remains below $10^{-10}$ throughout the simulation time.

As before, the energy separation between the two minima is given by $\Delta = 4.68$. Because of exponential suppression of the wave packet across the potential barrier, we find that detectable tunneling only occurs at energies sufficiently above $\Delta$, within the finite simulation times considered here. Especially for detecting tunneling from the right to the left well, maintaining higher energies becomes crucial. Accordingly, we set the energy baseline to $\Delta$ so that we can accurately compare the left-to-right and right-to-left tunneling. In the reduced dynamical system, this baseline is implemented as an energy offset applied to the right well, which amounts to a shift of the reference energy and ensures a consistent comparison between the two wells. Similar to the dynamical system studied before, here we continue to use $\langle x \rangle (t)$ as the primary indicator for detecting tunneling.

\begin{figure*}[!htbp]
\includegraphics[width=\textwidth]{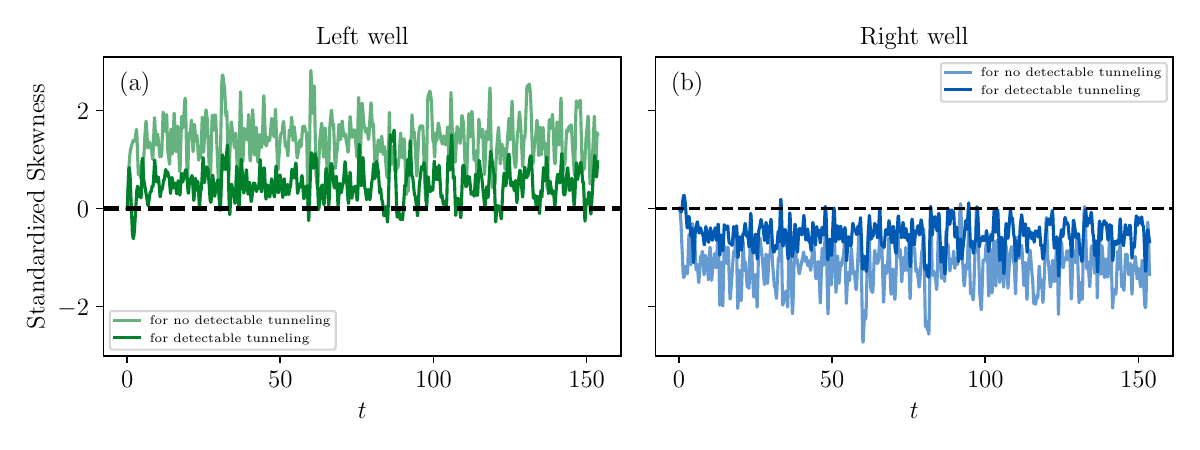}    
\caption{\textbf{Time evolution of standardized skewness: (a) left-well trajectories (left panel, green shades) and (b) right-well trajectories (right panel, blue shades)}. Light and dark shades indicate the two energies $E=9.0+\Delta$ (no tunneling) and $E=14.95+\Delta$ (tunneling), respectively, where $\Delta=4.68$ is the energy offset given by the depth of the right well relative to the left. Large skewness fluctuations correspond to regimes where tunneling is not detectable, while smaller fluctuations are associated with detectable tunneling between wells. Near-zero values of skewness indicate small asymmetry in the wave packet profile.}
\label{fig:skewness}
\end{figure*}

To ensure consistency with the dynamical systems analysis for the no-tunneling scenario, Gaussian wave packets are initialized at $x(0)=0.5$ and $x(0)=5.5$ in the left and right wells, respectively, both with total energy $E=9.0+\Delta$, where $\Delta$ denotes the energy offset between the minima of the wells. By tuning the wave packet's variance, we achieve energies that are nearly identical to those used in the dynamical systems simulations. The subsequent quantum dynamics, characterized by the time evolution of the mean position and variance, are presented in Fig.~\ref{fig:Schrodinger_no_tunnel-E-9}~(a) and (c) for the left well and Fig.~\ref{fig:Schrodinger_no_tunnel-E-9}~(b) and (d) for the right well, respectively.  The mean position fluctuates rapidly but stays within its well set by the initial location, indicating no tunneling.

Next, Gaussian wave packets are initialized at $x(0)=0.5$ and $x(0)=5.5$ in the left and right wells, respectively, both with total energy $E=14.95+\Delta$, again by adjusting the variance. At this energy, the mean position \(\langle x \rangle\) is seen to cross the barrier, indicating bi-directional tunneling from left to right (Fig.~\ref{fig:Schrodinger_tunnel-E-14-95}(a)) and from right to left (Fig.~\ref{fig:Schrodinger_tunnel-E-14-95}(b)). Although the amplitude of oscillations in \(\langle x \rangle\) is smaller than in Fig.~\ref{fig:dyn_sys_tunnel-E-14-95}(a) and (b), tunneling still takes place.

Moreover, the variance for the right well is similar to that in the left well in the no-tunneling case, as shown in Figs.~\ref{fig:Schrodinger_no_tunnel-E-9}(c) and (d). This suggests that while the wave packet's mean position is determined by the well, its fluctuations and spreading dynamics are governed by similar initial conditions and total energy. For the tunneling scenario, the corresponding time evolution of the variance \(V(x)\) is presented in the lower panels (c) and (d) of Fig.~\ref{fig:Schrodinger_tunnel-E-14-95}. The variances for both initializations are again similar in magnitude, though they exhibit larger fluctuations than in the non-tunneling case (Fig.~\ref{fig:Schrodinger_no_tunnel-E-9}(c) and (d)). This enhanced spreading underscores the role of total energy in governing wave packet dynamics, a feature that persists across both localized and tunneling regimes.

Quantitative differences between tunneling behavior obtained from the reduced dynamical systems and the full set of time-dependent Schr\"odinger equations (TDSE) are expected due to the truncation of fifth- and higher-order moments and the use of the Gaussian approximation for the kurtosis in the closure scheme. In particular, the reduced model tends to overestimate the excursions of $\langle x\rangle$, whereas the full TDSE captures wave packet splitting and interference effects during interaction with the barrier. Nevertheless, when the full spatial extent of the wave packet, including its standard deviation, is considered, both the reduced model and the TDSE results consistently indicate the onset of tunneling. Thus, despite these approximations, the reduced dynamical system captures the qualitative features of detectable tunneling, providing a useful estimate of the corresponding energy threshold.

\begin{figure*}[!htbp]
\includegraphics[width=\textwidth]{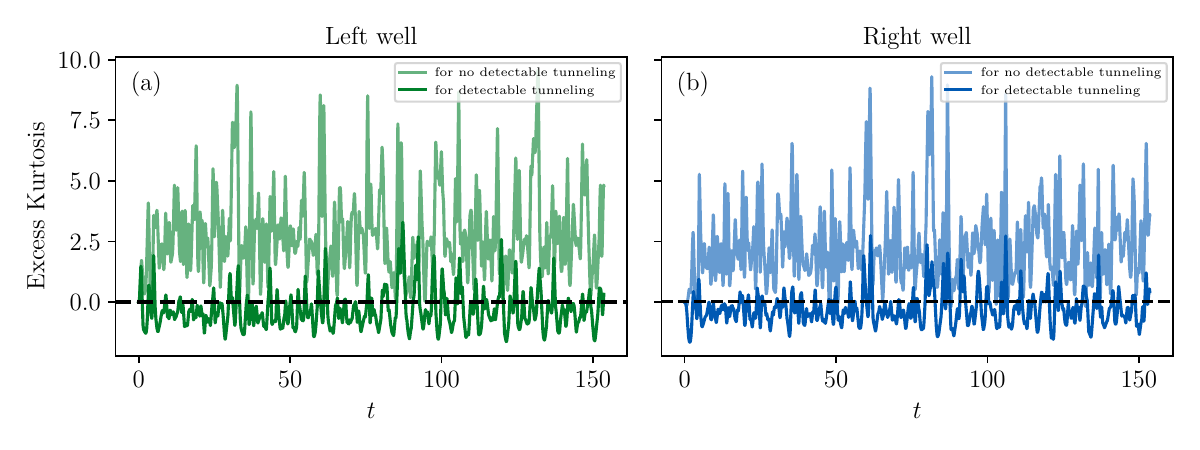}

\caption{\textbf{Time evolution of excess kurtosis: (a) left-well trajectories (left panel, green shades) and (b) right-well trajectories (right panel, blue shades)}. Light and dark shades indicate the two energies $E=9.0+\Delta$ (no tunneling) and $E=14.95+\Delta$ (tunneling), respectively, where $\Delta=4.68$ is the energy offset given by the depth of the right well relative to the left. Large kurtosis fluctuations correspond to regimes where tunneling is not detectable, while smaller fluctuations are associated with detectable tunneling between wells. Near-zero values of excess kurtosis indicate weak departures from a Gaussian profile of the wave packet.}

\label{fig:kurtosis}
\end{figure*}

To assess the reliability of the Gaussian moment-closure approximation underlying the Ehrenfest approach, we computed the time evolution of the wave packet's standardized skewness and excess kurtosis from TDSE,
\begin{align*}
\gamma_1(t)=\frac{\langle (x-\langle x\rangle)^3\rangle}{\langle (x-\langle x\rangle)^2\rangle^{3/2}},
\quad
\gamma_2(t)=\frac{\langle (x-\langle x\rangle)^4\rangle}{\langle (x-\langle x\rangle)^2\rangle^{2}}-3.
\end{align*}
These quantities measure, respectively, the asymmetry of the distribution and its deviation from a Gaussian profile, for which both are zero. As shown in Figs.~\ref{fig:skewness} and \ref{fig:kurtosis}, they remain bounded throughout the evolution for both no tunneling and tunneling cases. Although large transient deviations from zero occur, particularly before the detectable tunneling takes place, no runaway growth of these moments is observed in either case. This indicates that the distribution undergoes only moderate and transient non-Gaussian deformations during the evolution, supporting the validity of the closure scheme within the studied parameter regime.

The evolution of these higher moments also reveal additional structures in the dynamics. The skewness largely maintains a definite sign determined by the initial well: it remains positive for left-well initialization and negative for right-well initialization, independent of whether tunneling occurs. In both cases its magnitude is largest prior to the onset of observable tunneling and decreases afterward, leading to a less-asymmetric distribution. The excess kurtosis exhibits a similar transient behavior, with larger values before the onset of detectable tunneling and smaller values afterward. This reduction occurs irrespective of the well in which the wave packet is initialized. The observation suggests that the strongest non-Gaussian deformation of the wave packet develops when reflections from the barrier dominate, while the subsequent emergence of detectable tunneling is accompanied by smaller standardized skewness and kurtosis, indicating a recovery of an almost Gaussian form.

\section{Conclusion}

In conclusion, this work has demonstrated that a dynamical systems approach, rooted in the Ehrenfest theorem, provides a useful framework for analyzing detectable quantum tunneling in asymmetric double-well potentials. By approximating the quantum dynamics through a four-dimensional system of equations for the first- and second-order moments, the mean position $\langle x \rangle$, the variance $V$, and their time derivatives, we can track the system's evolution without directly solving the Schr\"odinger equation. Beyond the time-independent WKB approach, this time-dependent analysis also conserves the system's mean energy, which serves as a control parameter for the system, throughout the dynamics.

For tunneling in an asymmetric double-well potential, a crucial quantification of the asymmetry is provided by the third-order moment, skewness, of the wave packet. Although we neglect fifth or higher-order moments in this analysis (with a Gaussian approximation to treat the fourth-order kurtosis), the skewness, intrinsically set by the potential's parameters, directly encodes its directional asymmetry. It appears explicitly in the dynamical equations, influencing the evolution of both mean position and variance. 

From a dynamical systems perspective, the potential barrier corresponds to a classically unstable fixed point. Our stability analysis reveals an energy threshold beyond which it becomes stable, leading to continuous switching of the mean position $\langle x\rangle$ between the two wells. We identify this switching behavior as detectable tunneling across the potential barrier. Full numerical simulations of the time-dependent Schr\"odinger equation using an initial Gaussian wave packet confirm that this theoretically predicted threshold closely approximates the energy value above which detectable tunneling occurs, with small deviations attributed to the approximations inherent in our moment-truncation scheme.

In summary, the Ehrenfest theorem--based dynamical systems framework offers a robust and versatile theoretical framework for investigating bi-directional tunneling in isolated quantum systems. The present formulation can be readily extended to more complex, realistic potentials, such as multi-well structures relevant to molecular conformers and superconducting circuits. Furthermore, incorporating non-Gaussian initial states and developing refined moment-closure schemes within this framework could enable accurate analysis of anharmonic systems relevant to anharmonic vibrational spectroscopy. Together, these features highlight the dynamical systems approach as a practical and predictive tool for modeling quantum transport.

\subsection*{Acknowledgements}
The authors thank Jayanta K. Bhattacharjee for comments.

\section*{Author Declarations}

\subsection*{Conflict of Interest}
The authors have no conflicts to disclose.

\subsection*{Author contributions}
\textbf{Swetamber Das}  Conceptualization (equal); Formal analysis (supporting); Software (lead);  Visualization (lead); Writing -- original draft (equal); Writing -- review and editing (equal). 
\textbf{Arghya Dutta}: Conceptualization (equal); Formal analysis (lead); Software (supporting); Visualization (supporting); Writing -- original draft (equal); Writing -- review and editing (equal).

\subsection*{Data Availability}
Data sharing is not applicable to this article as no new data were created or analyzed in this study. All numerical codes supporting the findings of this study are openly available in the \href{https://github.com/Dynamics-and-Complexity-Group/quantum-dynamics}{GitHub repository}. A permanent versioned archive is also maintained on \href{https://doi.org/10.5281/zenodo.19426286}{Zenodo}.

\bibliography{ref}
\bibliographystyle{elsarticle-num}
\end{document}